\documentclass[conference]{IEEEtran}
\usepackage{cite}
\usepackage{amsmath,amssymb,amsfonts}
\usepackage{algorithmic}
\usepackage{graphicx}
\usepackage{tikz}
\usepackage{subcaption}
\usetikzlibrary{shapes,arrows}
\usepackage{multicol,lipsum}
\usepackage{lipsum}
\usepackage{mathtools}
\usepackage{cuted}
\usepackage{hyperref}
\usepackage{textcomp}
\usepackage{amsmath}
\usepackage{esvect}
\usepackage{bm}

\usepackage{amsmath,bbm,amssymb,amsfonts,amstext,amsopn}

\DeclareMathOperator{\Tr}{Tr}
\DeclareMathOperator{\Rank}{Rank}

\usepackage{lipsum,amsmath,multicol}
\usepackage{amsthm}
\usepackage{amssymb}
\usepackage{epstopdf}

\usepackage{amsthm}
\usepackage{cite}
\usepackage{balance}
\allowbreak
\usepackage{url}
\usepackage{epsfig}
\usepackage{setspace}
\usepackage{stmaryrd}
\usepackage{psfrag}	
\usepackage{multirow}
\usepackage{float}
\usepackage{amsthm}

\newcommand{\subparagraph}{}
\usepackage[compact]{titlesec}
\usepackage[english]{babel}
\usepackage[process=auto]{pstool}
\usepackage{amsthm}
\theoremstyle{remark}

\usepackage{etoolbox}
\usepackage{algorithm}
\usepackage[process=auto]{pstool}
\usepackage{epsfig}
\usepackage{caption}
\usepackage{algorithmic}
\usepackage{xcolor}
\newtheorem{thm}{Theorem}
\newtheorem{l em}{Lemma}
\newtheorem{prop}[thm]{Proposition}

\allowdisplaybreaks
\begin{document}
\title{Joint Beamforming and Phase Shift Optimization for Multicell IRS-aided OFDMA-URLLC Systems} 
\author{ Walid R. Ghanem,
Vahid Jamali, and Robert Schober  \\
Friedrich-Alexander-University Erlangen-Nuremberg, Germany \quad}\vspace{-1cm}
\maketitle
\begin{abstract}
	This paper investigates the resource allocation algorithm design for intelligent reflecting surface (IRS) aided multiple-input single-output (MISO) orthogonal frequency division multiple access (OFDMA) multicell networks, where a set of base stations cooperate to serve a set of ultra-reliable low-latency communication (URLLC) users. The IRS is deployed to enhance the communication channel and increase reliability by creating a virtual line of sight for URLLC users with unfavorable propagation conditions. This is the first study on IRS-enhanced OFDMA-URLLC systems. The resource allocation algorithm design is formulated as an optimization problem for the maximization of the weighted system sum throughput while guaranteeing the quality of service of the URLLC users. The optimization problem is non-convex and finding the globally optimal solution entails a high computational complexity which is not desirable for real-time applications. Therefore, a suboptimal iterative algorithm is proposed which \textit{jointly} optimizes all optimization variables in each iteration using a new iterative rank minimization approach. The algorithm is guaranteed to converge to a locally optimal solution of the formulated optimization problem. Our simulation results show that the proposed IRS design facilitates URLLC and yields large performance gains compared to two baseline schemes.
\end{abstract}
\section{Introduction}
Ultra-reliable low-latency communication (URLLC) is one of three important use cases of the fifth generation (5G) of wireless communication systems. URLLC focuses on mission-critical applications including factory automation, autonomous driving, remote surgery, and smart grid automation. These applications require sub-millisecond latency and very high reliability\cite{Mehdi1}. These strict quality-of-service (QoS) requirements are challenging to achieve, and thus, new technologies and system designs are needed.

 A promising approach to facilitate URLLC is the deployment of intelligent reflecting surfaces (IRSs). IRSs comprise a set of passive elements which  can reflect the incident signals with desired phase shifts\cite{Quirs1,marirsj}. By optimizing the IRS phase shifts, wireless channels can be proactively manipulated, and virtual line-of-sight (LOS) channels to the URLLC users can be created. Unfortunately, the existing resource allocation designs for IRS-assisted systems are not directly applicable to URLLC systems. The authors in \cite{ofdmairs} investigated resource allocation for orthogonal frequency division multiple access (OFDMA) IRS-aided systems. In \cite{alex1}, robust and secure resource allocation for IRS-assisted systems was proposed. Resource allocation for IRS-aided multicell networks was investigated in \cite{hua2020intelligent,irscell,celldai}. However, the studies in \cite{ofdmairs,alex1,hua2020intelligent,irscell,celldai} were based on Shannon's capacity formula for the additive white Gaussian noise (AWGN) channel. Since URLLC systems employ a short frame structure and short packet transmission (SPT) to reduce latency, the relation between the achievable rate, decoding error probability, and transmission delay cannot be captured by Shannon's capacity formula which assumes infinite block length and zero error probability \cite{shannon}. If Shannon's capacity formula is utilized for resource allocation design for IRS-aided URLLC systems, the latency will be underestimated and the reliability will be overestimated, and as a result, the QoS requirements of the users will not be met.  On the other hand, existing resource allocation schemes for SPT in URLLC systems, such as \cite{gha,chsecross,convexfinite}, do not exploit IRSs. Thus, the benefits of IRSs for facilitating URLLC has not been investigated, yet.
  
Furthermore, most resource allocation algorithms proposed for IRS-aided systems, such as \cite{Quirs1,irsswipt}, are based on alternating optimization and semi-definite relaxation (SDR) techniques. However, the combination of alternating optimization and SDR  methods does not ensure the local optimality of these algorithms. In particular, the Gaussian randomization required to meet the rank constraint after SDR is an impediment to the local optimality of the overall resource allocation algorithm. Thus, efficient new approaches for resource allocation design for IRS-assisted systems are needed. 

In this paper, we address the above issues and focus on the resource allocation algorithm design for multicell multiuser IRS-aided multiple-input single-output (MISO) OFDMA-URLLC systems. This paper makes the following main contributions:  
\begin{itemize}
	\item We tackle the resource allocation algorithm design for multicell multiuser IRS-aided MISO OFDMA-URLLC systems. The resource allocation algorithm design is formulated as an optimization problem with the objective to maximize the weighted sum throughput subject to QoS constraints for all URLLC users and unit modulus constraints for the IRS elements.
	\item The formulated problem is non-convex, e.g., due to the coupling between the variables and the structure of the expression for the achievable rate for SPT, and hence finding the global optimal solution entails a high computational complexity. To overcome this issue, we first transform the problem into a difference-of-convex (D.C.) programming problem. Subsequently, Taylor series approximation and successive convex approximation are employed to find a locally optimal solution of the original problem. Moreover, a novel iterative rank minimization method is proposed to guarantee the rank constraint after SDR. 
	\item We show by simulation that the proposed IRS-aided system facilitates URLLC and yields large performance gains compared to two baseline schemes.  
\end{itemize}

\textit{Notation}: Lower-case letters refer to scalar numbers, while bold lower and upper case letters denote vectors and matrices, respectively. $\Tr{(\mathbf{A})}$ and $\Rank{(\mathbf{A})}$ denote the trace and the rank of matrix $\mathbf{A}$, respectively. $\mathbf{A}\succeq 0$  indicates that matrix $\mathbf{A}$ is positive semi-definite. $\mathbf{A}^{H}$ and ${\mathbf{A}}^{T}$ denote the Hermitian transpose and the transpose of matrix $\mathbf{A}$, respectively. $\mathbb{C}$ is the set of complex numbers. $\mathbf{1}_{n}$ denotes the $n \times 1$ all-ones vector.  $\mathbf{I}_{N}$ is the $N \times N$ identity matrix. $\mathbb{H}_{N}$  denotes the set of all $N \times N$ Hermitian matrices. $|\cdot|$ and  $\|\cdot\|$ refer to the absolute value of a complex scalar and the Euclidean vector norm, respectively. The circularly symmetric complex Gaussian distribution with mean $\mu$ and variance $\sigma^{2}$ is denoted by $\mathcal{CN}(\mu,\sigma^{2})$, and $\sim$ stands for ``distributed as". $\mathcal{E}\{\cdot\}$ denotes statistical expectation. $\nabla_{x}f(\mathbf{x})$ denotes the gradient vector of function $f(\mathbf{x})$ and its elements are the partial derivatives of $f(\mathbf{x})$. $\|\mathbf{A}\|_{F}$ is the Frobenius norm of matrix $\mathbf{A}$. $j=\sqrt{-1}$ denotes the imaginary
unit of a complex number. $\text{diag}(\mathbf{a})$
represents a diagonal matrix whose main diagonal elements
are extracted from vector $\mathbf{a}$. Diag($\mathbf{A}$) denotes a vector whose elements are extracted from the main diagonal elements of matrix $\mathbf{A}$. $\mathbf{A}^{*}$ denotes the optimal value of  optimization variable $\mathbf{A}$. 
\section{System and Channel Models}
In this section, we present the system and channel models of the considered multicell multiuser IRS-assisted MISO OFDMA-URLLC system.
\subsection{System Model}
\begin{figure}
	\centering
	\scalebox{0.35}{
		\pstool{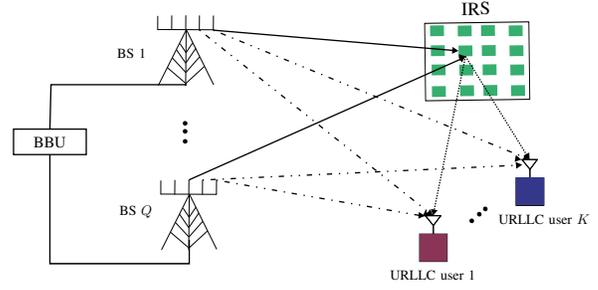}{
			\psfrag{a1}[c][c][1.5]{BS 1}
			\psfrag{aK}[c][c][1.5]{BS $Q$}
			\psfrag{u}[c][c][1.5]{\text{URLLC user 1}}
			\psfrag{u1}[c][c][1.5]{\text{URLLC user $K$}}
		    \psfrag{BBU}[c][c][1.7]{BBU}
		    \psfrag{irs}[c][c][2]{IRS}
	}}
	\caption{Multicell multiuser downlink URLLC system model with one BBU, $Q$ BSs, $K$ URLLC users, and one IRS.}
	\label{model}
	\vspace{-0.5cm}
\end{figure}

We consider a downlink system, where $Q$ BSs, indexed by $q =\{1,\dots,Q\}$ and equipped with $N_{T}$ antennas, serve $K$ single-antenna URLLC users, indexed by $k =\{1,\dots,K\}$, cf. Fig.~\ref{model}. A baseband processing unit (BBU) is deployed for control and planning. All BSs are connected to the BBU by optical cables
or wireless backhaul \cite{assumbbu}. The IRS is also controlled by the BBU or the BSs via wired or wireless links. The BSs simultaneously serve the URLLC users in the same frequency band. The frequency band is divided into $L$ subcarriers. An IRS is deployed to assist the communication between the BSs and the URLLC users. The IRS is equipped with a uniform planar array (UPA) composed of $M$ passive reflecting elements characterized by phase shift matrix $\boldsymbol{\Phi}=\text{diag}(\phi_{1}, \dots, \phi_{M})$,  where $\phi_{m}=e^{j\theta_{m}}$ and $\theta_{m}$ denotes the phase shift  of the $m$-th element of the IRS. A resource frame has a duration (the system delay) of $T_{f}$ seconds and contains $N_{p}=B_{s}T_{f}$ orthogonal frequency division multiplexing (OFDM) symbols, where $B_{s}$ is the subcarrier bandwidth. To obtain a performance upper bound, perfect channel state information (CSI) of the entire system is assumed to be available at the BBU for resource allocation design \cite{celldai} and the coherence time of the channels is assumed to exceed $T_{f}$, i.e., the channel is constant for all $N_{p}$ OFDM symbols of a frame. We assume that the delay requirements of all users are known at the BS and only users whose delay requirements can potentially be met in the current resource block are admitted into the system. The maximum transmit power of BS $q$ is $P_{q,\text{max}}$. Moreover, we assume the data of each user is available at all BSs. In addition, the BSs are synchronized and coordinate to perform beamforming for the users, i.e., they conduct coherent coordinate multipoint transmission and form a cell-free network\cite{hua2020intelligent,cell1,celldai}.   
 \subsection{Signal and Channel Models}
 In this paper, we assume each BS performs linear transmit precoding, where each user is assigned a unique beamforming vector. Hence, the transmit signal of BS $q$ on subcarrier $l$ is given by:
 \begin{IEEEeqnarray}{lll}\label{txvector}
 	\mathbf{x}_{q}[l]=\sum_{k=1}^{K}\mathbf{w}_{q,k}[l]u_{k}[l],
 \end{IEEEeqnarray}
 where $u_{k}[l] \in \mathbb{C}$ and $\mathbf{w}_{q,k}[l] \in \mathbb{C}^{N_{T} \times 1}$ are the transmit symbol and the beamforming vector of user $k$ on subcarrier $l$ at BS $q$, respectively. Moreover, without loss of generality, we assume  $\mathcal{E}\{|u_{k}[l]|^{2}\}=1, \; \forall k$. 
The received signal at user $k$ on subcarrier $l$ is  given as follows:
\begin{align}\label{receviedsignal}
	y_{k}[l]=\sum_{q=1}^{Q}(\mathbf{h}_{k}^{H}{[l]}\boldsymbol{\Phi}\mathbf{F}_{q}[l] +\mathbf{g}_{q,k}^{H}{[l]})\mathbf{x}_{q}[l]+z_{k}[l],
\end{align}
where $\mathbf{h}_{k}[l] \in \mathbb{C}^{M \times 1}$ denotes the channel vector from the IRS to user $k$ on subcarrier $l$, $\mathbf{F}_{q}[l] \in \mathbb{C}^{M \times N_{T}}$ is the channel matrix from BS $q$ to the IRS on subcarrier $l$, and $\mathbf{g}_{q,k}{[l]} \in \mathbb{C}^{N_{T}\times 1}$ is the channel vector from BS $q$ to user $k$  on subcarrier $l$. $z_{k}[l]\sim \mathcal{CN}(0,\sigma^{2})$ is the complex AWGN at user $k$ on subcarrier $l$. 
By substituting (\ref{txvector}) into (\ref{receviedsignal}), we obtain (\ref{receviedsignal3}), shown at the top of the next page. The signal-to-interference-plus-noise ratio (SINR) of user $k$ on subcarrier $l$ is given in (\ref{rho6}), shown at the top of the next page. 
\begin{figure*}\vspace{-0.75cm}
\begin{align}\label{receviedsignal3}\hspace{-0.25cm}
y_{k}[l]=\underbrace{\sum_{q=1}^{Q}(\mathbf{h}_{k}^{H}{[l]}\boldsymbol{\Phi}\mathbf{F}_{q}[l] +\mathbf{g}_{q,k}^{H}{[l]})\mathbf{w}_{q,k}[l]u_{k}[l]}_{\substack {\text {Desired signal of user $k$}}}\hspace{0.02cm}+\underbrace{\sum_{q=1}^{Q}\sum_{j\neq k}^{K}(\mathbf{h}_{k}^{H}{[l]}\boldsymbol{\Phi}\mathbf{F}_{q}[l] +\mathbf{g}_{q,k}^{H}{[l]})\mathbf{w}_{q,j}[l]u_{j}[l]}_{\substack {\text {multiuser interference (MUI)}}}+z_{k}[l],
\end{align}\vspace{-0.5cm}
\begin{align}{\label{rho6}}
\gamma_{k}[l]=\frac{\sum_{q=1}^{Q}|(\mathbf{h}_{k}^{H}{[l]}\boldsymbol{\Phi}\mathbf{F}_{q}[l] +\mathbf{g}_{q,k}^{H}{[l]})\mathbf{w}_{q,k}[l]|^{2}}{\sum_{q=1}^{Q}\sum_{j\neq k}^{K}|(\mathbf{h}_{k}^{H}{[l]}\boldsymbol{\Phi}\mathbf{F}_{q}[l] +\mathbf{g}_{q,k}^{H}{[l]})\mathbf{w}_{q,j}[l]|^{2}+\sigma^{2}}\triangleq\frac{r_{k}[l]}{d_{k}[l]}.
\end{align}\vspace{-0.5cm}
\end{figure*}
\section{Resource Allocation Problem Formulation}
In this section, we discuss the achievable rate for SPT, the QoS requirements of the URLLC users, and the adopted system performance metric for resource allocation algorithm design. Furthermore, we formulate the proposed resource allocation optimization problem for multicell multiuser IRS-aided MISO OFDMA-URLLC systems. 
\subsection{Achievable Rate for SPT}

For performance evaluation of SPT, the so-called normal approximation for finite blocklength codes was developed in \cite{thesis}. For parallel complex AWGN channels, the maximum number of bits $R$ conveyed in a packet comprising $L_{p}$ symbols can be approximated as\cite[Eq. (4.277)]{thesis},\cite[Fig. 1]{Erseghe1}:
\begin{IEEEeqnarray}{lll}\label{normalapproximation}
	R=\sum_{l=1}^{L_{p}}\log_{2}(1+\gamma[l])-Q^{-1}(\epsilon)\big(\sum_{l=1}^{L_{p}}{\nu}[l]\big)^{\frac{1}{2}},
\end{IEEEeqnarray}
where $\epsilon$ is the decoding packet error probability and $Q^{-1}(\cdot)$ is the inverse of the Gaussian Q-function with $
Q(x)=\frac{1}{\sqrt{2\pi}}\int_{x}^{\infty}\text{exp}{\left(-\frac{t^{2}}{2}\right)}\text{d}t$. Furthermore, 
\begin{IEEEeqnarray}{lll}\label{dispersion}
	\nu[l]=a^{2}\big(1-{(1+\gamma[l])^{-2}}\big)
\end{IEEEeqnarray} 
and $\gamma[l]$ are the channel dispersion \cite{thesis} and the SINR of the $l$-th symbol, respectively, where  $a=\log_{2}(\text{e})$.\color{black}

In this paper, we base the resource allocation algorithm design for multicell multiuser downlink IRS-aided MISO OFDMA-URLLC systems on (\ref{normalapproximation}). 
\subsection{QoS and System Performance Metric}
The QoS requirements of URLLC user $k$ include the minimum number of received bits, $B_{k}$, the target packet error probability, $\epsilon_{k}$, and the total system delay,  $T_{f}$. According to (\ref{normalapproximation}), the total number of bits transmitted over the resources allocated to user $k$ can be written as:
\begin{align}\label{BT}
	\bar{B}_{k}(\mathbf{w},\boldsymbol{\Phi})=C_{k}(\mathbf{w},\boldsymbol{\Phi})-V_{k}(\mathbf{w},\boldsymbol{\Phi}),
\end{align}
where\vspace{-0.85cm}
\begin{align}&
C_{k}(\mathbf{w}, \boldsymbol{\Phi})=N_{p}\sum_{l=1}^{L}\log_{2}(1+\gamma_{k}[l]),\\&
V_{k}(\mathbf{w},\boldsymbol{\Phi})=Q^{-1}(\epsilon_{k})\big({N_{p}\sum_{l=1}^L   V_{k}[l]}\big)^{\frac{1}{2}},
\end{align}
where the channel dispersion $V_{k}[l]$ is given by:
\begin{align}\label{disper}
V_{k}[l]=a^{2}\bigg(1-{(1+{\gamma_{k}}[l])}^{-2}\bigg).
\end{align}
Here, $\mathbf{w}$ is used to denote the collection of all beamforming  vectors $\mathbf{w}_{q,k}[l]$, $\forall q,k,l$. 
\subsection{Optimization Problem Formulation}
In the following, we formulate the resource allocation optimization problem for the maximization of the weighted sum throughput of the system subject to the QoS requirements of
each user regarding the received number of bits, the reliability,
and the latency as well as the unit modulus constraint of the  IRS elements. In particular, the proposed resource allocation policy is determined by solving the following optimization
problem:\vspace{-0.25cm}
\begin{align}\label{op1} &  \underset { {\mathbf {w},\mathbf{\Phi}}}{ \mathop {\mathrm {maximize}}\nolimits }~ \sum_{k=1}^K \mu_{k}(C_{k}(\mathbf{w},\mathbf{\Phi})-V_{k}(\mathbf{w},\mathbf{\Phi})) \\& \ \mbox {s.t.}~\mbox {C1}: \nonumber C_{k}(\mathbf{w},\mathbf{\Phi})-V_{k}(\mathbf{w},\mathbf{\Phi}) \geq B_{k},  \forall k, \\&\qquad \nonumber\mbox {C2}: \sum_{k=1}^K \sum_{l=1}^{L}\|\mathbf{w}_{q,k}[l]\|^{2}\leq P_{q,\text{max}},\forall q, \\&\qquad \nonumber  \mbox {C3}: |\phi_{m,m}|=1, \forall m,  
 \end{align}
where $\mu_{k}$ is the weight assigned to user $k$. Larger values of $\mu_{k}$ give a user a higher priority, and as a result, a higher throughput (i.e., more bits are transmitted to that user) compared to the other users. The values of the $\mu_{k}$ can be specified in the medium access control (MAC) layer and are assumed to be given in the following.
 
In (\ref{op1}), constraint $\mbox {C1}$ guarantees the transmission of a minimum number of $B_{k}$ bits to user $k$. Constraint $\mbox {C2}$ is the total power budget constraint of BS $q$. Finally, constraint $\mbox {C3}$ is the unit modulus constraint for the IRS elements. The problem in (\ref{op1}) is  non-convex. The non-convexity is caused by the non-convex form of the SINR in (\ref{rho6}), the coupling of the optimization variables, and the non-convex normal approximation in (\ref{BT}), which appears in the cost function and in constraint $\mbox {C1}$.

There is no systematic approach for solving general non-convex optimization problems. However, in the next section, we show that based on a series of transformations, problem (\ref{op1}) can be reformulated as a D.C. programming problem. This reformulation allows the application of Taylor series approximation to obtain a local optimum solution with low-computational complexity.
\section{Solution of the Optimization Problem}
In this section, we focus on solving the optimization problem formulated in (\ref{op1}) and obtain a locally optimal solution. We apply three different transformations to problem (\ref{op1}) to arrive at a D.C. optimization problem. Subsequently, SCA and a novel iterative rank minimization method are employed to solve the resulting problem.
The main steps for deriving the proposed algorithm are summarized in Fig.~\ref{alg}.
 \begin{figure*}[h!]	
	\centering
	\tikzstyle{decision} = [diamond, draw, fill=red!12, 
	text width=5em, text badly centered, node distance=3.5cm, inner sep=0pt]
	\tikzstyle{block} = [rectangle, draw, fill=red!12, 
	text width=2.5cm, text centered, rounded corners, minimum height=1.5cm]
   \tikzstyle{block2} = [rectangle, draw, fill=green!12, 
	text width=2.5cm, text centered, rounded corners, minimum height=1.5cm]
	\tikzstyle{line} = [draw, -latex']
	\tikzstyle{opt} = [rectangle,text badly centered,draw,fill=orange!20, node distance=2.5cm, minimum height=1.7cm]
	\scalebox{.7}{
		\begin{tikzpicture} [node distance =3.2cm, auto] 
		\node[opt,node distance=8cm]  (a) {Non-convex Problem (\ref{op1})};  
		\node[block,right of= a,node distance=4.2cm] (b) {1- Monotonic Reformulation}; 
		\node[block,right of=b] (c) {2- Semi-definite Programming Reformulation};
		\node[block,right of=c] (f) {3- D.C. Reformulation};
		\node[block,right of=f] (e) {4- SCA and Iterative Rank Minimization};
		\node[block2,right of=e] (h) {5- Locally Optimal Solution};
		\draw[line] (a)-- (b);  
		\draw[line] (b)-- (c);  
		\draw[line] (c)-- (f);
		\draw[line] (f)-- (e);  
		\draw[line] (e)-- (h);  
		\end{tikzpicture}} 
	\caption{The main steps for solving non-convex problem (\ref{op1}).}
	\label{alg}\vspace{-0.5cm}
\end{figure*}
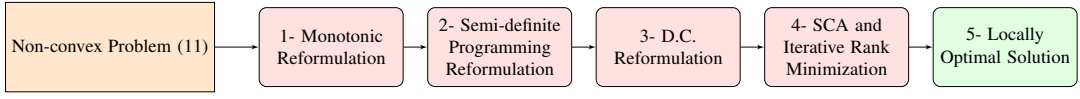\\
\subsection{Problem Transformation 1: Monotonic Reformulation}
In this subsection, we transform optimization problem (\ref{op1}) into an equivalent but more tractable form.
The objective function and constraint ${\mbox{C1}}$ in (\ref{op1}) have a complicated structure. In fact, the objective function in (\ref{op1}) and constraint ${\mbox{C1}}$ are differences of two functions that are monotonic in the SINR, i.e., $\gamma_{k}[l]$. To handle the related complexity and to facilitate the design of an efficient solution for problem (\ref{op1}), we introduce a set of auxiliary variables  $\chi_{k}[l],\forall k,l$, to bound the SINR from below, i.e.,
\begin{align}{\label{rho3}}
0  \leq \chi_{k}[l]\leq \gamma_{k}[l] \triangleq\frac{r_{k}[l]}{d_{k}[l]}, \forall k,l,
\end{align} 
where $	r_{k}[l]$ and $d_{k}[l]$ are the numerator and denominator of the SINR in (\ref{rho6}), respectively. Let us replace $\gamma_{k}[l]$ by $\chi_{k}[l]$ in $C(\mathbf{w},\mathbf{\Phi})$, $V(\mathbf{w},\mathbf{\Phi})$, $C_{k}(\mathbf{w},\mathbf{\Phi})$, and $V_{k}(\mathbf{w},\mathbf{\Phi})$ and denote the resulting functions by  $C(\boldsymbol{\chi})$, $V(\boldsymbol{\chi})$, $C_{k}(\boldsymbol{\chi}_{k})$, and $V_{k}(\boldsymbol{\chi}_{k})$, respectively, i.e.,
\begin{align} 
C({\boldsymbol{\chi}})=\sum_{k=1}^{K}\mu_{k}C_{k}(\boldsymbol{\chi}_{k}), \; \qquad
V({\boldsymbol{\chi}})=\sum_{k=1}^{K}\mu_{k}V(\boldsymbol{\chi}_{k}),
\end{align}\vspace{-0.4cm}
\begin{align}
\hspace{-2.2cm}	C_{k}(\boldsymbol{\chi}_{k})=N_{p}\sum_{l=1}^{L}\log_{2}(1+\chi_{k}[l]),
\end{align} \vspace{-0.4cm}
\begin{align}\hspace{-0.4cm}
V_{k}(\boldsymbol{\chi}_{k})=aQ^{-1}(\epsilon_{k})\big({N_{p}\sum_{l=1}^{L}(1-(1+\chi_{k}[l])^{-2})}\big)^{\frac{1}{2}}\;,
\end{align} 
where $\boldsymbol{\chi}_{k}$ denotes the collection of optimization variables $\chi_{k}[l], \ \forall l$, and $\boldsymbol{\chi}$ denotes the collection of optimization variables $\boldsymbol{\chi}_{k}, \forall k$. Using these notations, we formulate a new optimization problem as follows: 	  
\begin{align}\label{opequivalnt}&  \underset { {{\mathbf {w}}}, \boldsymbol{\Phi}, \boldsymbol {\chi}}{ \mathop {\mathrm {maximize}}\nolimits }~ C(\boldsymbol {\chi})-V(\boldsymbol {\chi}) \\& \ \mbox {s.t.}~ \widetilde{ \mbox {C1}}:\ \nonumber C_{k}({\boldsymbol{\chi}}_{k})-V_{k}(\boldsymbol{\chi}_{k}) \geq B_{k},  \forall k, \;\;\nonumber {\mbox {C2}},{\mbox {C3}},
\\& \nonumber \qquad \mbox {C4} : \chi_{k}[l]\leq \frac{r_{k}[l]}{d_{k}[l]}, \forall k,l, \;\;
\nonumber  \mbox {C5} : \chi_{k}[l] \geq 0. \;
\end{align} 
Unfortunately, the objective function and constraint $\widetilde{ \mbox {C1}}$ of problem (\ref{opequivalnt}) are not monotonic in $\boldsymbol {\chi}$, and thus, constraint $\mbox {C4}$ may not hold with equality. To cope with this issue, in the following, we transform (\ref{opequivalnt}) into a monotonic optimization problem in $\boldsymbol {\chi}$. Subsequently, we show that problem (\ref{op1}) and the reformulated monotonic problem are equivalent. To this end, recall that the main condition required for a problem to be a monotonic optimization problem is the monotonicity of the objective function and the constraints. We note that the objective function and constraint $\widetilde{ \mbox {C1}}$ in (\ref{opequivalnt}) are differences of two monotonic concave functions in the optimization variables $\boldsymbol{\chi}$\cite{gha}. Hence, problem (\ref{opequivalnt}) can be transformed into the canonical form of a monotonic optimization problem in two steps. 
\begin{itemize}
	\item \textbf{Step 1:} First, we transform the objective function in (\ref{opequivalnt}) into a monotonic function. Since  $V(\boldsymbol {\chi})$ is monotonically increasing in $\boldsymbol {\chi}$, $\boldsymbol {\chi} \leq \boldsymbol {\chi}_{\text{max}}$ leads to $V(\boldsymbol {\chi}) \leq V(\boldsymbol {\chi}_{\text{max}})$. Therefore, $V(\boldsymbol {\chi})+t = V(\boldsymbol {\chi}_{\text{max}})$ holds, for some positive $t$. Hence, substituting $V(\boldsymbol {\chi})$ by $V(\boldsymbol {\chi}_{\text{max}})-t$, the optimization problem in (\ref{opequivalnt}) can be rewritten as follows:   	
	\begin{align}\label{optimization4bb1}& \underset {{{\mathbf {w}}}, \boldsymbol{\Phi}, \boldsymbol {\chi}, t}{ \mathop {\mathrm {maximize}}\nolimits }~ C(\boldsymbol{\chi})+t-V(\boldsymbol {\chi}_{\text{max}}) \\& \ \mbox {s.t.}~ \nonumber \ \widetilde{ \mbox {C1}}, {\mbox {C2}},{\mbox {C3}}, \mbox{C4, C5},
	\\& \qquad ~\mbox {C6}:\nonumber t +V(\boldsymbol {\chi}) \leq V({\boldsymbol {\chi}_{\text{max}}}), \;\; ~\mbox {C7}: \nonumber t \geq 0,
	\end{align} 
	where the value of $V(\boldsymbol {\chi}_{\text{max}})$ is obtained by substituting ${\boldsymbol {\chi}}_{\text{max}}$ into $V(\boldsymbol {\chi})$ and  $\boldsymbol {\chi}_{{\text{max}}}$ is the collection of all  $\chi_{{\text{max}},k}[l], \forall k,l$, which can be obtained by removing the MUI in (\ref{rho6}) and by allocating all available power of each BS $q$ to subcarrier $l$. Hence, this corresponds to the high signal-to-noise ratio (SNR) regime, and since we are interested in $V(\boldsymbol {\chi}_{\text{max}})$ and not in $\boldsymbol {\chi}_{\text{max}}$ itself, the value of $V(\boldsymbol {\chi}_{\text{max}})$ can be obtained by considering the limits of (\ref{disper}) for high SINR. This leads to $V(\boldsymbol {\chi}_{\text{max}})=\sum_{k=1}^{K}\mu_{k}aQ^{-1}(\epsilon_{k})\sqrt{LN_{p}}$.
	
	\item \textbf{Step 2:} We use a similar approach as for transforming the cost function to transform  constraint $\widetilde{ \mbox {C1}}$ into a standard monotonic constraint. In particular, $V_{k}(\boldsymbol {\chi}_{k}) +\zeta_{k} = V_{k}(\boldsymbol {\chi}_{{\text{max}},k})$ holds for some positive auxiliary optimization variable $\zeta_{k}$, where $\boldsymbol {\chi}_{{\text{max}},k}$ is the collection of all $\chi_{{\text{max}},k}[l], \forall l$. Therefore, by substituting $V_{k}(\boldsymbol {\chi}_{k})$ by $  V_{k}(\boldsymbol {\chi}_{{\text{max},k}}) -\zeta_{k}$, constraint $\widetilde{\mbox {C1}}$ can be transformed into two monotonic constraints as follows:
	\begin{align}&\label{optimization4bs1}
	\widetilde{\mbox {C1a}}: \quad C_{k}(\boldsymbol {\chi}_{k}) + \zeta_{k} \geq V_{k}(\boldsymbol {\chi}_{{\text{max},k}})+B_{k},\forall k, \\&\label{optimization4bs2}
	\widetilde{\mbox {C1b}}: \quad V_{k}(\boldsymbol {\chi}_{k}) +\zeta_{k} \leq V_{k}(\boldsymbol {\chi}_{{\text{max},k}}),\forall k.
	\end{align}
\end{itemize}
We note that, at the optimal point, constraint $\mbox{C6}$ holds with equality due to the monotonicity of the objective function with respect to auxiliary optimization variable $t$. Moreover, the left hand sides of (\ref{optimization4bs1}) and (\ref{optimization4bs2}) are monotonically increasing functions. Hence, problem (\ref{optimization4bb1}) can be transformed into an equivalent  monotonic optimization problem in $\boldsymbol {\chi}$ as follows\cite{Tuy}:
\begin{align}\label{optimization22b}& \underset {{{\mathbf {w}}}, \boldsymbol{\Phi}, \boldsymbol {\chi},{t}, \boldsymbol{{\zeta}}}{ \mathop {\mathrm {maximize}}\nolimits }~ C(\boldsymbol {\chi})+t\\&  \ \mbox {s.t.}~  \nonumber \widetilde{\mbox {C1a}}, \widetilde{\mbox {C1b}}, \mbox {C2-C7}, 
\end{align}
where $\boldsymbol{\zeta}$ is the collection of optimization variables $\zeta_{k},\forall k$. Note that, in (\ref{optimization22b}),  we removed the constant $V(\boldsymbol {\chi}_{\text{max}})$ from the objective function because it has no effect on the optimal solution. Optimization problem (\ref{optimization22b}) has a monotonically increasing objective function in $\boldsymbol {\chi}$ and all constraints involving $\boldsymbol {\chi}$ are monotonically increasing functions. Now, due to the monotonicity of the objective function, $\widetilde{\mbox {C1a}}$, and $\widetilde{\mbox {C1b}}$ with respect to $\boldsymbol {\chi}$, constraint $\mbox{C4}$ is active at the optimal solution. Therefore, the inequality in $\mbox{C4}$ has to hold with equality which implies that optimization problem (\ref{optimization22b}) is equivalent to problem (\ref{op1}). In the following, we focus on finding a locally optimal solution for transformed problem (\ref{optimization22b}).
\subsection{Problem Transformation 2: SDP Programming Reformulation}
In the following, we transform (\ref{optimization22b}) into an equivalent SDP problem. To this end, we exploit the following identity\cite{alexpower} 
\begin{align}{\label{tk}}&\hspace{-2cm}
|(\mathbf{h}_{k}^{H}{[l]}\boldsymbol{\Phi}\mathbf{F}_{q}[l] +\mathbf{g}_{q,k}^{H}{[l]})\mathbf{w}_{q,k}[l]|^{2}\nonumber\\&=\mathbf{v}^{H}\mathbf{R}_{q,k}^{H}[l]\mathbf{W}_{q,k}[l]\mathbf{R}_{q,k}[l]\mathbf{v},
\end{align}
where $\mathbf{R}_{q,k}[l]=[\mathbf{F}_{q}^{H}[l]\text{diag}(\mathbf{h}_{k}^{H}[l])\;\; \mathbf{g}_{k}[l]]$, $\mathbf{W}_{q,k}[l]=\mathbf{w}_{q,k}[l]\mathbf{w}_{q,k}^{H}[l]$, $\mathbf{v}=[\tilde{\mathbf{v}}\;\; x]$, $|x|^{2}=1$, and $\tilde{\mathbf{v}}=[e^{j\theta_{1}},e^{j\theta_{2}},\dots,e^{j\theta_{M}}]$. 
 Let us define slack optimization variables $I_{k}[l], \forall k, l,$ to upper bound the denominator of $\gamma_{k}[l], \forall k,l,$ in constraint $\mathrm{C4}$, and rewrite constraint $\mathrm{C4}$ equivalently as:  
\begin{align}&\label{qq1}
\mathrm{C4a}: {\chi}_{k}[l]I_{k}[l] \leq 
{r}_{k}[l],\forall k,l,\\&
\mathrm{C4b}:I_{k}[l] \geq d_{k}[l], \forall k,l,
\end{align} 
where $I_{k}[l]$ represents the interference plus the noise of user $k$ on subcarrier $l$.
Now, optimization problem (\ref{optimization22b}) can be rewritten as an equivalent SDP problem as follows:
\begin{align}\label{opf}\hspace{-0.5cm} &\underset {\mathbf {W}, \mathbf {V}, \boldsymbol {\chi},{t}, \boldsymbol{{\zeta}},\bar{\mathbf{I}}}{ \mathop {\mathrm {maximize}}\nolimits }~ C(\boldsymbol {\chi})+t\\&  \ \mbox {s.t.}~  \nonumber \widetilde{\mbox {C1a}}, \widetilde{\mbox {C1b}}, \; 
\overline{\mbox {C2}}:\sum_{k=1}^{K}\sum_{l=1}^{L}\Tr(\mathbf{W}_{q,k}[l])\leq {P}_{q,\text{max}}, \forall q,  
 \nonumber \\& \quad  \nonumber \mbox {C4a}:{\chi}_{k}[l]I_{k}[l] \leq  \sum_{q=1}^{Q}\Tr(\mathbf{W}_{q,k}[l]\mathbf{G}_{q,k}[l]\mathbf{V}\mathbf{G}_{q,k}^{H}[l]), \forall k,l,
 \nonumber \\& \quad \mbox {C4b}: \sum_{q=1}^{Q}\sum_{j \neq k}^{K}\Tr(\mathbf{W}_{q,j}[l]\mathbf{G}_{q,k}[l]\mathbf{V}\mathbf{G}_{q,k}^{H}[l])+\sigma^{2}\leq I_{k}[l], \forall k,l,
\nonumber  \\& \quad \mathrm{C5,C6},\mathrm{C7}, \mathrm{C8}:\Rank(\mathbf{V})=1, \mbox {C9}: \mathbf{V} \succeq	0,\;\nonumber  \\& \quad\mbox {C10}:\text{Diag}(\mathbf{V})=\mathbf{1}_{M+1},\nonumber
 \quad \mbox {C11}: \Rank(\mathbf{W}_{q,k}[l])=1, \forall q,k,l,\nonumber \\& \quad \mbox {C12}: \mathbf{\mathbf{W}}_{q,k}[l] \succeq	0,\forall q,k,l,\nonumber
\end{align} 
where $\bar{\mathbf{I}}$ denotes the collection of variables $I_{k}[l], \forall k,l$.
We note that $\Rank({{\mathbf{W}}_{q,k}}[l]) \leq 1,  \forall q,k,l,$ and ${\mathbf{W}}_{q,k}[l]\succeq 0$ in constraints $\mbox{C11}$ and $\mbox{C12}$ are imposed to ensure that  $\mathbf{W}_{q,k}[l]=\mathbf{w}_{q,k}[l]\mathbf{w}_{q,k}^{H}[l]$ holds after optimization. Similarly, $\Rank({{\mathbf{V}}}) \leq 1$ and ${\mathbf{V}}\succeq 0$ in constraints $\mbox{C8}$ and $\mbox{C9}$ are imposed to ensure that  $\mathbf{V}=\mathbf{v}\mathbf{v}^{H}$ holds after optimization.  Moreover, for simplicity of notation, we define ${\mathbf {W}}$ as the collection of all Hermitian matrices $\mathbf{W}_{q,k}[l] \in \mathbb{H}_{N_{T}}$,~$\forall q,k,l$.  
\subsection{Problem Transformation 3: D.C. Programming Reformulation}
In the following, the goal is to reformulate problem (\ref{opf}) in the canonical form of D.C. programming. This will be done in two main steps as follows:  

\textbf{Step 1:} The bilinear term ${\chi}_{k}[l]I_{k}[l] $ on the left hand side of $\mathrm{C4a}$ is non-convex and an obstacle to designing a computationally efficient resource allocation algorithm. 
However, this product can be written as the difference of two convex functions as follows\cite{Tuybookgo}:
\begin{align}&\label{qq2b}
\hspace{-1.4cm}\overline{\mathrm{C4a}}: f_{1}(\chi_{k}[l],I_{k}[l])-f_{2}(\chi_{k}[l],I_{k}[l]) \leq \\&\qquad
\sum_{q=1}^{Q}\Tr(\mathbf{W}_{q,k}[l]\mathbf{G}_{q,k}[l]\mathbf{V}\mathbf{G}_{q,k}^{H}[l]), \forall k,l,\nonumber
\end{align}
where 
\begin{align}&\label{qq2c}
f_{1}(\chi_{k}[l],I_{k}[l])= 0.5({\chi}_{k}[l]+I_{k}[l])^{2}\\&
f_{2}(\chi_{k}[l],I_{k}[l])=0.5({\chi}_{k}[l])^{2}+0.5(I_{k}[l])^{2}.
\end{align}
\textbf{Step 2:} Constraints $\overline{\mathrm{C4a}}$ and $\mathrm{C4b}$ are non-convex due to the coupling between the beamforming and the phase shift matrices. To tackle this issue, we apply difference of convex decomposition to these constraints. In particular, for two Hermitian matrices $\mathbf{A}$ and $\mathbf{B}$, the following relation holds for the Frobenius inner product: 
\begin{align}\hspace{-0.25cm}\label{dc2a}
\Tr(\mathbf{A}\mathbf{B})=\frac{1}{2}\|\mathbf{A}+\mathbf{B}\|^{2}_{F}-\frac{1}{2}\|\mathbf{A}\|^{2}_{F}-\frac{1}{2}\|\mathbf{B}\|^{2}_{F}.
\end{align}
Eq.~(\ref{dc2a}) is in the form of a difference of two convex functions. Using (\ref{dc2a}), we can rewrite constraints $\overline{\mathrm{C4a}}$ and $\mbox{C4b}$ in the following difference-of-convex functions form:
\begin{align}&\hspace{-1cm}\label{c6aa}
\overline{\overline{\mbox{C4a}}}:0.5\sum_{q=1}^{Q}\bigg(f_{3}(\mathbf{W}_{q,k}[l],\mathbf{V})-f_{4}(\mathbf{W}_{q,k}[l],\mathbf{V})\bigg)\geq \nonumber \\&\hspace{1cm} f_{1}(\chi_{k}[l],I_{k}[l])-f_{2}(\chi_{k}[l],I_{k}[l]), \forall k,l,
\end{align}
where
\begin{align}&\label{c6a2}\hspace{-0.2cm}
f_{3}(\mathbf{W}_{q,k}[l],\mathbf{V})=\|\mathbf{W}_{q,k}[l]+\mathbf{R}_{q,k}[l]\mathbf{V}\mathbf{R}_{q,k}^{H}[l]\|^{2}_{F}\\&\hspace{-0.2cm}
f_{4}(\mathbf{W}_{q,k}[l],\mathbf{V})=\|\mathbf{W}_{q,k}[l]\|^{2}_{F}+\|\mathbf{R}_{q,k}[l]\mathbf{V}\mathbf{R}_{q,k}^{H}[l]\|^{2}_{F}, 
\end{align}
and
\begin{align}&\hspace{-0.2cm}\label{6c}
\overline{\mbox{C4b}}:0.5\sum_{q=1}^{Q}\sum_{j \neq k}^{K}\bigg(f_{5}(\mathbf{W}_{q,j}[l],\mathbf{V})-f_{6}(\mathbf{W}_{q,j}[l],\mathbf{V})\bigg)+\sigma^{2}\nonumber\\&\hspace{5cm}\leq I_{k}[l], \forall j\neq k,l,
\end{align}
where
\begin{align}&\label{c6bb}\hspace{-0.65cm}
f_{5}(\mathbf{W}_{q,j}[l],\mathbf{V})=\|\mathbf{W}_{q,j}[l]+\mathbf{R}_{q,k}[l]\mathbf{V}\mathbf{R}_{q,k}^{H}[l]\|^{2}_{F},\\&\hspace{-0.65cm}
f_{6}(\mathbf{W}_{q,j}[l],\mathbf{V})=\|\mathbf{W}_{q,j}[l]\|^{2}_{F}+\|\mathbf{R}_{q,k}[l]\mathbf{V}\mathbf{R}_{q,k}^{H}[l]\|^{2}_{F}.
\end{align}
We have transformed the coupling in constraints $\overline{\mbox{C4a}}$ and $\mbox{C4b}$ into differences of two convex functions constraints  $\overline{\overline{\mbox{C4a}}}$ and $\overline{\mbox{C4b}}$, respectively. Next, we use Taylor series approximations to obtain convex approximations for non-convex constraints $\overline{\overline{\mbox{C4a}}}$ and $\overline{\mbox{C4b}}$. Then, based on these approximations, we provide the proposed iterative algorithm that finds a locally optimal solution of the original optimization problem (\ref{op1}).
\subsection{SCA and Iterative Rank Minimization Approach (IRMA) }
In this subsection, we present the proposed iterative algorithm. We apply Taylor series approximation and iterative rank minimization, which finally leads to an SCA algorithm. 

\textbf{Step 1 (Taylor series approximation):} To obtain a convex optimization problem that can be efficiently solved, we have to handle the non-convex D.C. constraints $\widetilde{\mbox {C1b}}$, $\mbox {C6}$, $\overline{\overline{\mbox{C4a}}}$, and $\overline{\mbox{C4b}}$ using Taylor series approximations for functions ${V}_{k}(\boldsymbol{\chi}_{k})$, $f_{2}(\chi_{k}[l],I_{k}[l])$, $f_{3}(\mathbf{W}_{q,k}[l],\mathbf{V})$, and $f_{6}(\mathbf{W}_{q,k}[l],\mathbf{V})$. For function ${V}_{k}(\boldsymbol{\chi}_{k})$, for feasible points $\boldsymbol{\chi}_{k}^{({i})}, \forall k,$ where $i$ is the SCA iteration index, the following inequality holds:  
 \begin{align}&\hspace{-0.4cm}
\label{inequalit1b}
{V}_{k}(\boldsymbol{\chi}_{k}) \leq \overline{V}_{k}(\boldsymbol{\chi}_{k}) =  {V}_{k}(\boldsymbol{\chi}_{k}^{({i})})+\nabla_{\boldsymbol{\chi}_{k}}{V}_{k}(\boldsymbol{\chi}_{k}^{({i})})^{T}(\boldsymbol{\chi}_{k}-\boldsymbol{\chi}_{k}^{({i})}),
\end{align}
where $\nabla_{\boldsymbol{\chi}_{k}}{V}_{k}(\boldsymbol{\chi}_{k}^{({i})})$ is the gradient of ${V}_{k}(\boldsymbol{\chi}_{k})$. Moreover, the Taylor series approximations of functions 
$f_{2}(\chi_{k}[l],I_{k}[l])$, $f_{3}(\mathbf{W}_{q,k}[l],\mathbf{V})$, and $f_{6}(\mathbf{W}_{q,j}[l],\mathbf{V})$, denoted by
 $\bar{f}_{2}(\chi_{k}[l],\chi^{(i)}_{k}[l],z^{(i)}_{k}[l],z_{k}[l])$ and $\bar{f}_{d}(\mathbf{W}_{q,k}[l],\mathbf{V},\mathbf{W}^{(i)}_{q,k}[l],\mathbf{V}^{(i)}), \forall d=\{3,6\}$, respectively, are given as follows:
	\begin{align}&\label{f1bar}\hspace{-0.6cm}
\bar{f}_{2}(\chi_{k}[l],\chi^{(i)}_{k}[l],z^{(i)}_{k}[l],z_{k}[l])\nonumber\\&\hspace{-0.25cm}=0.5({\chi}^{(i)}_{k}[l])^{2}+{\chi}^{(i)}_{k}[l]({\chi}_{k}[l]-{\chi}^{(i)}_{k}[l])\nonumber\\&\hspace{0.5cm}+0.5({z}^{(i)}_{k}[l])^{2}+{z}^{(i)}_{k}[l]({z}_{k}[l]-{z}^{(i)}_{k}[l]), \forall k,l,
\end{align}
and
\begin{align}&\hspace{-0.15cm}\label{TS3}
\bar{f}_{d}(\mathbf{W}_{q,k}[l],\mathbf{V})\geq  \bar{f}_{d}(\mathbf{W}_{q,k}[l],\mathbf{V},\mathbf{W}^{(i)}_{q,k}[l],\mathbf{V}^{(i)})\nonumber\\&+\Tr\big(\nabla_{\mathbf{W}}\bar{f}_{d}(\mathbf{W}^{(i)}_{q,k}[l],\mathbf{V}^{(i)})^{T}(\mathbf{W}_{q,k}[l]-\mathbf{W}^{(i)}_{q,k}[l])\big)\nonumber\\&\hspace{-0.2cm}+\Tr\big(\nabla_{\mathbf{V}}\bar{f}_{d}(\mathbf{W}^{(i)}_{q,k}[l],\mathbf{V}^{(i)})^{T}(\mathbf{V}-\mathbf{V}^{(i)})\big), \forall d=\{3,6\},
\end{align}
where $\nabla_{\mathbf{W}}\bar{f}_{d}(\mathbf{W}^{(i)}_{q,k}[l],\mathbf{V}^{(i)})$ and $\nabla_{\mathbf{V}}\bar{f}_{d}(\mathbf{W}^{(i)}_{q,k}[l],\mathbf{V}^{(i)}$ are the gradients of $\bar{f}_{d}(\mathbf{W}_{q,k}[l],\mathbf{V})$ with respect to $\mathbf{W}$ and $\mathbf{V}$, respectively. By substituting (\ref{inequalit1b})-(\ref{TS3}) into (\ref{opf}), we obtain the following optimization problem:
\begin{align}\label{opfaa}& \underset {{{\mathbf {W}}}, \boldsymbol{V}, \boldsymbol {\chi},{t}, \boldsymbol{{\zeta}},\bar{\mathbf{I}}}{ \mathop {\mathrm {maximize}}\nolimits }~ C(\boldsymbol {\chi})+t\\&  \ \mbox {s.t.}~  \nonumber \widetilde{\mbox {C1a}},\quad  \overline{\mbox {C1b}}:\quad \overline{V}_{k}(\boldsymbol {\chi}_{k}) +\zeta_{k} \leq V_{k}(\boldsymbol {\chi}_{{\text{max},k}}),\forall k.
\\&\nonumber \overline{\overline{\overline{\mbox {C4a}}}}:f_{1}(\chi_{k}[l],I_{k}[l])-
\bar{f}_{2}(\chi_{k}[l],z_{k}[l])\nonumber \leq \\&\hspace{3cm}
\bar{f}_{3}(\mathbf{W}_{q,k}[l],\mathbf{V})-f_{4}(\mathbf{W}_{q,k}[l],\mathbf{V}), \forall k,l,\nonumber
\\&\nonumber \overline{\overline{\mbox{C4b}}}:f_{5}(\mathbf{W}_{q,k}[l],\mathbf{V})-\bar{f}_{6}(\mathbf{W}_{q,k}[l],\mathbf{V})+\sigma^{2}\leq I_{k}[l], \forall k,l,\\&\nonumber \mbox{C5},\; \widetilde{\mbox{C6}}: \overline{V}(\boldsymbol {\chi}) +t \leq V_{k}(\boldsymbol {\chi}_{{\text{max}}}), \mbox{C7-C12}.
\end{align}
The remaining difficulties in solving (\ref{opfaa}) are the non-convex rank constraints $\mbox {C8}$ and $\mbox {C11}$, which are tackled in the following.   

\textbf{Step 2 (IRMA)}:
To handle constraint $\mbox {C8}$, we exploit the following proposition: 
\begin{prop}[see \cite{sun2016iterative2}]\label{pro1}
A non-zero positive semi-definite matrix, $\mathbf{A}\in \mathbb{H}_{n}$, is a rank-one matrix if and only if $r\mathbf{I}_{n-1}-\mathbf{Y}^{T}\mathbf{A}\mathbf{Y}\succeq 0$ holds for $r=0$, where $\mathbf{Y} \in \mathbb{R}^{n \times (n-1)}$ is a matrix whose columns are those eigenvectors of $\mathbf{A}$ which correspond to the $n-1$ smallest eigenvalues.
\end{prop} 
Using \textbf{Proposition} \ref{pro1}, we substitute rank constraint $\mbox {C8}$ by semi-definite constraint $r\mathbf{I}_{n-1}-\mathbf{Y}^{T}\mathbf{V}\mathbf{Y}\succeq 0$, and penalize the value of $r$ in the objective function such that eventually $r=0$ and  $\Rank(\mathbf{V})=1$ are satisfied. Hence, optimization problem (\ref{opfaa}) is rewritten as follows:
\begin{align}\label{opfa}& \underset {{{\mathbf {W}}}, \boldsymbol{V}, \boldsymbol {\chi},{t}, \boldsymbol{{\zeta}}, \bar{\mathbf{I}}, r^{(i)}}{ \mathop {\mathrm {maximize}}\nolimits }~ C(\boldsymbol {\chi})+t-\eta_{1}^{(i)}r^{(i)}\\&  \ \mbox {s.t.}~  \nonumber \widetilde{\mbox {C1a}}, \overline{\mbox {C1b}}, \overline{\mbox {C2}},\overline{\overline{\overline{\mbox {C4a}}}}, \overline{\overline{\mbox {C4b}}}, \mbox{C5}, \widetilde{\mbox{C6}}, \mbox{C7,C9-C12}, \\&\qquad
\overline{\mbox{C8}}:r^{(i)}\mathbf{I}_{n-1}-\mathbf{Y}_{i-1}^{T}\mathbf{V}\mathbf{Y}_{i-1}\succeq 0,\nonumber
\end{align}
where $\eta_{1}^{(i)}$ is a weighting factor for $r^{(i)}$ in the $i$-iteration, $\mathbf{Y}_{i-1}$ are the eigenvectors corresponding to the $n-1$ smallest eigenvalues of the $\mathbf{V}$ obtained in the previous iteration $i-1$. In each iteration $i$, we try to maximize the weighed sum system throughput and at the same time minimize the newly introduced auxiliary optimization variable $r^{(i)}$ by increasing the weighting factor $\eta_{1}^{(i)}$, such that eventually $r^{(i)} = 0$ holds, and thus, the rank-one constraint on $\mathbf{V}$ is satisfied after convergence. Moreover, for constraint $\mbox {C11}$, we employ well-known SDR. The tightness of SDR for the problem at hand can be proved following the same steps
as in \cite{gha,alexpower}.  Due to the space constraints, we omit
the detailed proof. Problem (\ref{opfa}) after SDR of constraint $\mbox {C11}$ becomes a convex optimization problem  which can be solved using standard optimization software tools such as CVX \cite{cvx}. 
\begin{algorithm}[t]	
	\begin{algorithmic}[1]\label{sca2} 
	\caption{Successive Convex Approximation and IRMA}
	\STATE {Initialize:} Generate random initial points $\mathbf{W}^{(1)}$, $\mathbf{V}^{(1)}$, $\boldsymbol {\chi}^{(1)}$, $\mathbf {z}^{(1)}$. Set iteration index $i=1$, maximum number of iterations $I_{\text{max}}$,  initial penalty factors, $\eta_{1}^{(1)} >0$, $\eta_{2}^{(1)} >0$, maximum value of penalty factors  $\eta_{1,\text{max}}$ and $\eta_{2,\text{max}}$, $\alpha_{1}$, and $\alpha_{2}$. \\
	\STATE \textbf{Repeat}\\
	\STATE Solve convex problem (\ref{opfa2}) for given   $\mathbf{W}^{(i)}$, $\mathbf{V}^{(i)}$, $\boldsymbol {\chi}^{(i)}$, $\mathbf {z}^{(i)}$, $\mathbf{Y}^{(i)}$, and store the intermediate solutions   $\mathbf{W}$, $\mathbf{V}$, $\boldsymbol {\chi}$, $\mathbf {z}$\\
	\STATE Set ${i}={i}+1$ and update $\mathbf{W}^{(i)}=\mathbf{W}$, $\mathbf{V}^{(i)}=\mathbf{V}$,
	$\boldsymbol {\chi}^{(i)}=\boldsymbol {\chi}$,
	$\mathbf {z}^{(i)}$=$\mathbf {z}$,
	\STATE Find orthonormal eigenvectors $\mathbf{Y}^{(i)}$ of $\mathbf{V}^{(i)}$  
	\STATE Update $\eta_{1}^{(i)} =\min(\alpha_{1}\eta_{1}^{(i)},\eta_{1,\text{max}})$, $\eta_{2}^{(i)} =\min(\alpha_{2}\eta_{2}^{(i)},\eta_{2,\text{max}})$ 
	\STATE \textbf{Until} convergence or $i=I_{\text{max}}$.\\
	
\STATE {Output:} $\mathbf{W}^{*}=\mathbf{W}$,
	$\mathbf{V}^{*}=\mathbf{V}$.
	\end{algorithmic} 
\end{algorithm} 

\textbf{Algorithm Initialization:}
To solve optimization problem (\ref{opfa}), we need feasible initial points to achieve high performance. However, in general, it is not easy to find such initial points. To overcome this issue, we relax optimization problem (\ref{opfa}). To do so, we first rewrite all constraints in the form $g_{c} \leq 0, \forall c,$ where $c$ refers to a given constraint. Then, we relax these constraints as $g_{c} \leq \beta^{(i)}, \forall c,$ by adding a slack variable $\beta^{(i)}\geq0$. At the same time, we penalize the violation of these constraints by adding a penalty value $\eta_{2}^{(i)}$ for $\beta^{(i)}$ in the objective function of problem (\ref{opfa}) and increase the penalty value $\eta_{2}^{(i)}$ in each iteration $i$. Thus, problem (\ref{opfa}) can be rewritten as follows:
\begin{align}\label{opfa2}& \underset {\big({{\mathbf {W}}}, \boldsymbol{V}, \boldsymbol {\chi},{t}, \boldsymbol{{\zeta}}, \bar{\mathbf{I}}, r^{(i)}, \beta^{(i)}\big) \in \mathcal{Y}}{ \mathop {\mathrm {maximize}}\nolimits }~ C(\boldsymbol {\chi})+t-\eta_{1}^{(i)}r^{(i)}-\eta_{2}^{(i)}\beta^{(i)},
\end{align} 
where $\mathcal{Y}$ is the relaxed feasible set of problem (\ref{opfa}). \textbf{Algorithm} 1 presents an iterative algorithm for solving (\ref{opfa2}). In the first iteration, by choosing small penalty weights $\eta_{1}^{(1)} >0$ and $\eta_{2}^{(1)}> 0$, we allow the constraints to be violated such that the relaxed feasible set $\mathcal{Y}$ is large. Then, in each subsequent iteration $i$, we use the solution from the previous iteration as initial point, increase the penalty weights $\eta_{1}^{(i)}$ and $\eta_{2}^{(i)}$ via multiplication factors $\alpha_{1}>1$ and $\alpha_{2}>1$, respectively, and solve the problem
again. Thus, if a feasible point exists, for sufficiently large values of $\eta_{1,\text{max}}$ and $\eta_{2,\text{max}}$ \cite{eaxtpenalty,pccp},  continuing this iterative procedure eventually yields a solution where $r^{(i)}=0$ and $\beta^{(i)}=0$ hold. The maximum values $\eta_{1,\text{max}}$ and $\eta_{2,\text{max}}$ for the penalty weights are imposed to avoid numerical instability. Moreover, since problem (\ref{op1}) is reformulated as a D.C. problem and Taylor series approximation is used to convexify the problem, according to \cite{fastglobal,pccp},  \textbf{Algorithm} 1 produces a sequence of improved feasible solutions until convergence to a local optimum point of problem (\ref{opfaa}) or equivalently problem (\ref{op1}) in polynomial time\cite{sun2016iterative2,pccp,fastglobal}.  
\section{Performance Evaluation}
In this section, we provided simulation results to validate the performance of the considered system. We adopt the simulation parameters provided in Table I, unless specified otherwise. We assume the center of the network is the point $(0,0)~\textrm{m}$ and two BSs are located at $(0,-100)~\textrm{m}$ and $(0,100)~\textrm{m}$ while the IRS is located at $(50,0)$. Moreover, the users are located inside a circle with radius 5~$\textrm{m}$ and the center of the circle is the point $(25,0)~\textrm{m}$. Let $d_{BS,U}$, $d_{BS,IRS}$, and $d_{IRS,U}$ denote the distance between a BS and a user, a BS and the IRS, and the IRS and a user, respectively. The distance-dependent path loss of the BS-user link is given by
$
PL_{d}=h_{b}(\frac{\lambda}{4\pi})^{2}d^{-\kappa_{BS,U}}_{BS,U},
$
and that of the BS-IRS-user link is given by\footnote{Note that the BS-IRS-user link is affected by the double-fading effect discussed in\cite{backscatter}.} 
$
PL_{r}=\bar{h}_{b}(\frac{\lambda}{4\pi})^{4}d^{-\kappa_{BS,IRS}}_{BS,IRS}d^{-\kappa_{IRS,U}}_{IRS,U}$, where $\lambda$ is the wavelength and $h_{b}$ and $\bar{h}_{b}$  represent the large-scale fading coefficients that represent the shadowing/blockage in the direct channel and the BS-IRS-user channel, respectively\cite{marirsj,celldai}. The path loss exponents of the BS-user, BS-IRS, and IRS-user link, are set as $\kappa_{BS,U}=3.2$, $\kappa_{BS,IRS}=2.1$, and $\kappa_{IRS,U}=2.1$, respectively. Moreover, for the small-scale fading, we assume Rayleigh fading for the BS-user channel and Rician fading with Rician factor of $10$ for the BS-IRS and IRS-user channels, respectively. The system delay is $T_{f}=0.41667~\textrm{ms}$, i.e., $N_{p}=100$. The parameters of \textbf{Algorithm} 1 are set as $I_{\text{max}}=500$, $\eta_{1}^{(1)} =10^{-2}$, $\eta_{2}^{(1)} =10^{5}$, $\eta_{1,\text{max}}=10^{2}$, $\eta_{2,\text{max}}=10^{8}$, $\alpha_{1}=25$, and $\alpha_{2}=25$.
\begin{table}[t]
	\label{tab:table}
	\centering
	\caption{Simulation Parameters.} 
	\renewcommand{\arraystretch}{1.4}
	\scalebox{0.75}{%
		\begin{tabular}{|c||c|}
			\hline
			Parameter & Value \\ \hline \hline 
			Carrier center frequency & 6 \textrm{GHz} \\ \hline
			Total number  of subcarriers $L$ & 32 \\ \hline
			Bandwidth of each sub-carrier & 240 kHz \\ \hline
			Noise power density  & -174 dBm/Hz \\ \hline
			Maximum BS transmit power, $P_{q,\text{max}}$  &  $45$~dBm \\ \hline  			
		User weights
						 &  $\mu_{k}=1,\forall k,$						\\ \hline
	${h}_{b}$ and	$\bar{h}_{b}$ & -80, 0 \textrm{dB}					\\ \hline  
		Packet error probability & $\epsilon_{k}=10^{-6}, \forall k$ \\ \hline    
	 Number of bits per packet & $B_{k}=160$~\textrm{bits}, $\forall k$ \\ \hline  
	\end{tabular}}
\end{table} 
\subsection{Performance Bound and Benchmark Schemes}
We compare the performance of the proposed resource allocation algorithm with the following schemes:
\begin{itemize}
	\setlength{\itemsep}{1pt}
	\item {\textbf{Shannon's capacity (SC)}}: To obtain an (unachievable) upper bound on the system performance, Shannon's capacity formula is adopted in problem (\ref{op1}), i.e., $V_{k}(\mathbf{w},\mathbf{\Phi}),\forall k,$ is set to zero. The resulting optimization problem is solved using a modified version of the proposed algorithm. 
	\setlength{\itemsep}{1pt}
	\item {\textbf{Baseline 1}}: In this scheme, we adopt random phase shifts for the IRS elements and optimize the beamforming of the BSs.
	\item {\textbf{Baseline 2}}: In this scheme, we remove the IRS from the system model and optimize the beamforming of the BSs.
\end{itemize}
\subsection{Simulation Results}
\begin{figure}[t!]
	\centering
	\resizebox{0.85\linewidth}{!}{\psfragfig{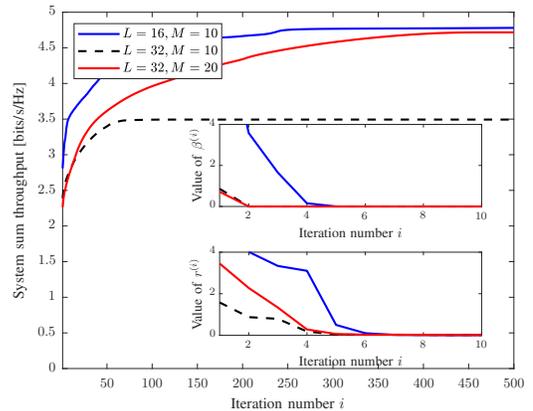}}\vspace{-4mm}
	\caption{Convergence of \textbf{Algorithm} 1. $Q=2$, $K=2$, $N_{T}=2$, $\forall k$, $P_{q,\text{max}}=45$ dBm, $\forall k$.} 
	\label{con}\vspace{-0.5cm}
\end{figure}

Fig.~\ref{con} shows the convergence of \textbf{Algorithm} 1 for a given channel realization. As can be observed, the system sum throughput increases monotonically with the number of iterations and reaches a local optimal point of the formulated optimization problem after convergence. Moreover, the values of $r^{(i)}$ and $\beta^{(i)}$ are also shown in Fig.~\ref{con}  as functions of the number of iterations. In particular, since we increase the penalty value $\eta_{1}^{(i)}$ in each iteration $i$, the value of $r^{(i)}$ decreases until it is equal to zero, such that the rank-one constraint on $\mathbf{V}$ is satisfied. The value of $\beta^{(i)}$ also decreases with the number of iterations and reaches zero after a few iterations, such that problem (\ref{opfa2}) becomes equivalent to problem (\ref{opfa}). 

\begin{figure}[t!]
	\centering
	\resizebox{0.85\linewidth}{!}{\psfragfig{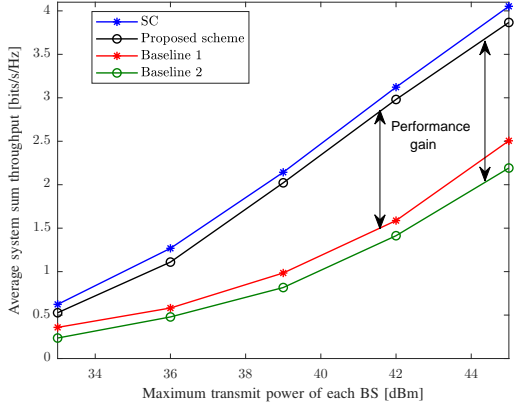}}\vspace{-4mm}
	\caption{The average system sum throughput versus the maximum transmit power of each BSs. $Q=2$, $L=32$, $N_{T}=4$, $M=20$, $K=2$, $\forall k$.} 
	\label{pow}\vspace{-0.7cm}
\end{figure}

Fig.~\ref{pow} shows the average system sum throughput versus the maximum total transmit power of the BSs, $P_{q,\text{max}}$. As can be observed from Fig.~\ref{pow}, the average system sum throughput increases monotonically with the maximum transmit power of the BSs. The reason for this is that the SINRs of the URLLC users can be improved by increasing the transmit power, which leads to an improvement of the system sum throughput. Fig.~\ref{pow} also shows that the average system sum throughput of the proposed scheme exceeds that of the considered baseline schemes. In particular, for Baseline 2, there is no IRS in the system, and thus, the average system sum throughput is limited by poor channel conditions between the users and the BSs. Moreover, the proposed scheme attains a large performance gain with respect to Baseline 1, as the latter applies random phase shifts for the IRS elements and does not utilize the benefits of passive IRS beamforming. In Fig.~\ref{pow}, we also compare the performance of the proposed scheme with SC. SC provides an upper bound for the average system sum throughput of the system. However, SC cannot guarantee the required latency and reliability. This is due to the fact that SC does not take into account the performance loss incurred by SPT for resource allocation design, and thus, the obtained resource allocation policies may not meet the QoS constraints. Nevertheless, the proposed scheme closely approaches the performance of SC.

\begin{figure}[t!]
	\centering
	\resizebox{0.85\linewidth}{!}{\psfragfig{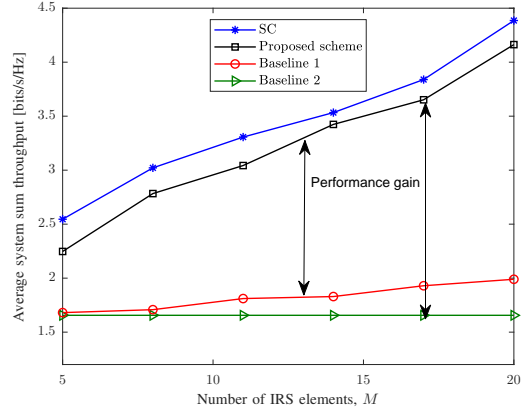}}\vspace{-4mm}
	\caption{The average system sum throughput versus the number of IRS elements. $Q=2$, $L=16$, $N_{T}=2$, $K=2$, $P_{\text{max}}=40$ dBm.} 
		\label{element}\vspace{-0.7cm}
\end{figure}

Fig.~\ref{element} shows the impact of the number of IRS elements $M$ on the average system sum throughput. As can be observed from Fig.~\ref{element}, the average system sum throughput increases as the number of elements increase for all considered schemes employing an IRS.  The reason for this is that more IRS passive reflecting elements $M$ can reflect more of the signal power received from the BS which leads to a larger power gain.  The achievable rate for Baseline 2 has a slower growth rate in terms of $M$ compared to the proposed scheme as it does not optimize the phase shifts.
\section{Conclusions} 
	This paper has studied the resource allocation algorithm design for multicell multiuser IRS-aided MISO OFDMA-URLLC systems, where a set of BSs cooperates to serve a set of URLLC users. The IRS was deployed to enhance the communication channel and to increase the reliability by providing a virtual LOS for the URLLC users. We formulated the resource allocation algorithm as a non-convex optimization problem for the maximization of the weighted system sum throughput subject to QoS constraints for the URLLC users. A low-complexity algorithm for jointly optimizing all variables by exploiting a new iterative rank minimization approach was proposed and shown to find a locally optimal point. Our simulation results have confirmed that an IRS can facilitate URLLC and the proposed algorithm achieves a large performance gain compared to two baseline scheme.

\bibliography{ref}  

\begin{thebibliography}{10}
\providecommand{\url}[1]{#1}
\csname url@samestyle\endcsname
\providecommand{\newblock}{\relax}
\providecommand{\bibinfo}[2]{#2}
\providecommand{\BIBentrySTDinterwordspacing}{\spaceskip=0pt\relax}
\providecommand{\BIBentryALTinterwordstretchfactor}{4}
\providecommand{\BIBentryALTinterwordspacing}{\spaceskip=\fontdimen2\font plus
\BIBentryALTinterwordstretchfactor\fontdimen3\font minus
  \fontdimen4\font\relax}
\providecommand{\BIBforeignlanguage}[2]{{%
\expandafter\ifx\csname l@#1\endcsname\relax
\typeout{** WARNING: IEEEtran.bst: No hyphenation pattern has been}%
\typeout{** loaded for the language `#1'. Using the pattern for}%
\typeout{** the default language instead.}%
\else
\language=\csname l@#1\endcsname
\fi
#2}}
\providecommand{\BIBdecl}{\relax}
\BIBdecl

\bibitem{Mehdi1}
M.~Bennis, M.~Debbah, and H.~V. Poor, ``Ultra reliable and low-latency wireless
  communication: Tail, risk, and scale,'' \emph{Proc. IEEE}, vol. 106, no.~10,
  pp. 1834--1853, Oct. 2018.

\bibitem{Quirs1}
Q.~{Wu} and R.~{Zhang}, ``Intelligent reflecting surface enhanced wireless
  network via joint active and passive beamforming,'' \emph{IEEE Trans. Wirel.
  Commun.}, vol.~18, no.~11, pp. 5394--5409, Aug. 2019.

\bibitem{marirsj}
M.~Najafi, V.~Jamali, R.~Schober, and H.V.Poor, ``Physics-based modeling and
  scalable optimization of large intelligent reflecting surfaces,''
  https://arxiv.org/abs/2004.12957, 2020.

\bibitem{ofdmairs}
Y.~{Yang}, S.~{Zhang}, and R.~{Zhang}, ``{IRS}-enhanced {OFDMA}: Joint resource
  allocation and passive beamforming optimization,'' \emph{IEEE Wireless
  Commun. Lett.}, vol.~9, no.~6, pp. 760--764, Jan. 2020.

\bibitem{alex1}
X.~{Yu}, D.~{Xu}, Y.~{Sun}, D.~W.~K. {Ng}, and R.~{Schober}, ``Robust and
  secure wireless communications via intelligent reflecting surfaces,''
  \emph{IEEE J. Sel. Areas Commun.}, pp. 1--1, July 2020.

\bibitem{hua2020intelligent}
M.~Hua, Q.~Wu, D.~W.~K. Ng, J.~Zhao, and L.~Yang, ``Intelligent reflecting
  surface-aided joint processing coordinated multipoint transmission,''
  https://arxiv.org/abs/2003.13909, 2020.

\bibitem{irscell}
C.~{Pan}, H.~{Ren}, K.~{Wang}, W.~{Xu}, M.~{Elkashlan}, A.~{Nallanathan}, and
  L.~{Hanzo}, ``Multicell {MIMO} communications relying on intelligent
  reflecting surfaces,'' \emph{IEEE Trans. Wirel. Commun.}, vol.~19, no.~8, pp.
  5218--5233, May 2020.

\bibitem{celldai}
Z.~Zhang and L.~Dai, ``A joint precoding framework for wideband reconfigurable
  intelligent surface-aided cell-free network,''
  https://arxiv.org/abs/2002.03744, 2020.

\bibitem{shannon}
C.~E. Shannon, ``A mathematical theory of communication,'' \emph{Bell Syst.
  Tech. J}, vol.~56, no.~5, pp. 2307--2359, May 2010.

\bibitem{gha}
W.~R. {Ghanem}, V.~{Jamali}, Y.~{Sun}, and R.~{Schober}, ``Resource allocation
  for multi-user downlink {MISO OFDMA-URLLC} systems,'' \emph{IEEE Trans.
  Commun.}, pp. 1--1, 2020.

\bibitem{chsecross}
C.~She, C.~Yang, and T.~Q.~S. Quek, ``Cross-layer optimization for
  ultra-reliable and low-latency radio access networks,'' \emph{{IEEE} Trans.
  Commun}, vol.~17, no.~1, pp. 127--141, Jan 2018.

\bibitem{convexfinite}
S.~Xu, T.~H. Chang, S.~C. Lin, C.~Shen, and G.~Zhu, ``Energy-efficient packet
  scheduling with finite blocklength codes: convexity analysis and efficient
  algorithms,'' \emph{{IEEE} Trans. Wireless Commun}, vol.~15, no.~8, pp.
  5527--5540, Aug 2016.

\bibitem{irsswipt}
Y.~{Tang}, G.~{Ma}, H.~{Xie}, J.~{Xu}, and X.~{Han}, ``Joint transmit and
  reflective beamforming design for {IRS}-assisted multiuser {MISO SWIPT}
  systems,'' in \emph{ICC 2020 - 2020 IEEE International Conference on
  Communications (ICC)}, 2020, pp. 1--6.

\bibitem{assumbbu}
U.~{Siddique}, H.~{Tabassum}, and E.~{Hossain}, ``Downlink spectrum allocation
  for in-band and out-band wireless backhauling of full-duplex small cells,''
  \emph{IEEE Trans. Commun.}, vol.~65, no.~8, pp. 3538--3554, April 2017.

\bibitem{cell1}
E.~{Nayebi}, A.~{Ashikhmin}, T.~L. {Marzetta}, H.~{Yang}, and B.~D. {Rao},
  ``Precoding and power optimization in cell-free massive {MIMO} systems,''
  \emph{IEEE Trans. Wirel. Commun.}, vol.~16, no.~7, pp. 4445--4459, May 2017.

\bibitem{thesis}
Y.~Polyanskiy, ``Channel coding: {N}on-asymptotic fundamental limits,'' Ph.D.
  dissertation, Princeton University.

\bibitem{Erseghe1}
T.~Erseghe, ``Coding in the finite-blocklength regime: {B}ounds based on
  {L}aplace integrals and their asymptotic approximations,'' \emph{IEEE Trans.
  Inf. Theory}, vol.~62, no.~12, pp. 6854--6883, Dec 2016.

\bibitem{Tuy}
\BIBentryALTinterwordspacing
H.~Tuy, F.~Al-Khayyal, and P.~T. Thach, \emph{{M}onotonic {O}ptimization:
  {B}ranch and {C}ut {M}ethods}.\hskip 1em plus 0.5em minus 0.4em\relax Boston,
  MA: Springer US, 2005, pp. 39--78. [Online]. Available:
  \url{https://doi.org/10.1007/0-387-25570-2_2}
\BIBentrySTDinterwordspacing

\bibitem{alexpower}
X.~Yu, D.~Xu, D.~W.~K. Ng, and R.~Schober, ``Power-efficient resource
  allocation for multiuser {MISO} systems via intelligent reflecting
  surfaces,'' https://arxiv.org/abs/2005.06703, 2020.

\bibitem{Tuybookgo}
H.~Tuy, \emph{Convex Analysis and Global Optimization}, 2nd~ed.\hskip 1em plus
  0.5em minus 0.4em\relax Springer Publishing Company, Incorporated, 2016.

\bibitem{sun2016iterative2}
C.~Sun and R.~Dai, ``An iterative rank penalty method for nonconvex
  quadratically constrained quadratic programs,'' \emph{SIAM Journal on Control
  and Optimization}, vol.~57, no.~6, pp. 3749--3766, 2019.

\bibitem{cvx}
M.~Grant and S.~Boyd, ``{CVX}: Matlab software for disciplined convex
  programming, version 2.1,'' \url{http://cvxr.com/cvx}, Mar. 2014.

\bibitem{eaxtpenalty}
H.~A. L.~T. andTao Pham~Dinh and H.~V. Ngai, ``Exact penalty and error bounds
  in {DC} programming,'' \emph{J. Glob. Optim.}, vol.~52, p. 509–535, Sep.
  2011.

\bibitem{pccp}
T.~Lipp and S.~Boyd, ``Variations and extension of the convex--concave
  procedure,'' \emph{Optim. Eng.}, vol.~17, no.~2, pp. 263--287, Jun 2016.

\bibitem{fastglobal}
H.~H. Kha, H.~D. Tuan, and H.~H. Nguyen, ``Fast global optimal power allocation
  in wireless networks by local {D.C.} programming,'' \emph{{IEEE} Trans.
  Wireless Commun}, vol.~11, no.~2, pp. 510--515, February 2012.

\bibitem{backscatter}
J.~D. {Griffin} and G.~D. {Durgin}, ``Complete link budgets for
  backscatter-radio and {RFID} systems,'' \emph{IEEE Antennas and Propagation
  Magazine}, vol.~51, no.~2, pp. 11--25, July 2009.

\end{thebibliography}


\begin{thebibliography}{}
\providecommand{\url}[1]{#1}
\csname url@samestyle\endcsname
\providecommand{\newblock}{\relax}
\providecommand{\bibinfo}[2]{#2}
\providecommand{\BIBentrySTDinterwordspacing}{\spaceskip=0pt\relax}
\providecommand{\BIBentryALTinterwordstretchfactor}{4}
\providecommand{\BIBentryALTinterwordspacing}{\spaceskip=\fontdimen2\font plus
\BIBentryALTinterwordstretchfactor\fontdimen3\font minus
  \fontdimen4\font\relax}
\providecommand{\BIBforeignlanguage}[2]{{%
\expandafter\ifx\csname l@#1\endcsname\relax
\typeout{** WARNING: IEEEtran.bst: No hyphenation pattern has been}%
\typeout{** loaded for the language `#1'. Using the pattern for}%
\typeout{** the default language instead.}%
\else
\language=\csname l@#1\endcsname
\fi
#2}}
\providecommand{\BIBdecl}{\relax}
\BIBdecl

\end{thebibliography}
\bibliographystyle{IEEEtran}
\end{document}